\documentclass[aps,showpacs,twocolumn, prb]{revtex4-1}
\usepackage{hyperref}
\usepackage{graphicx}
\usepackage{amsmath,amssymb,amsfonts}
\usepackage{upgreek}
\usepackage{booktabs}
\usepackage{indentfirst}

\begin{document}

\title{Superconductivity in the Superhard Boride WB$_{4.2}$}

\author{Elizabeth M. Carnicom$^{1,*}$}
\author{Judyta Strychalska-Nowak$^2$}
\author{P. Wi\'{s}niewski$^3$}
\author{D. Kaczorowski$^3$}
\author{W. W. Xie$^4$}
\author{T. Klimczuk$^2$}
\author{R. J. Cava$^{1,*}$}

\affiliation{$^1$Department of Chemistry, Princeton University, Princeton, New Jersey 08544.}
\affiliation{$^2$Department of Physics, Gdansk University of Technology, Gdansk Poland 80-233.}
\affiliation{$^3$Institute of Low Temperature and Structure Research, Polish Academy of Sciences,
PNr 1410, 50-950 Wrocław, Poland}
\affiliation{$^4$Department of Chemistry, Louisiana State University, Baton Rouge, LA 70803.}

\begin{abstract}

We show that the superhard boride WB$_{4.2}$ is a superconductor with a \textit{T}$_c$ of 2.05(5) K. Temperature-dependent magnetic susceptibility, electrical resistivity, and specific heat measurements were used to characterize the superconducting transition. The Sommerfeld constant $\gamma$ for WB$_{4.2}$ is 2.07(3) mJ mol$^{-1}$ K$^{-2}$ and the $\Delta$C/$\gamma$\textit{T}$_c$ = 1.56, which is somewhat higher than what is expected for weakly coupled BCS type superconductors. The \textit{H}$_{c2}$ vs \textit{T} plot is linear over a wide temperature range, but does show signs of flattening by the lowest temperatures studied and therefore the zero-temperature upper critical field ($\mu_0$\textit{H}$_{c2}$(0)) for WB$_{4.2}$ lies somewhere between the linear extrapolation of $\mu_0$\textit{H}$_{c2}$(\textit{T}) to 0 K and expectations based on the WHH model.  

\end{abstract}
\maketitle

\section{Introduction}

	High superconducting transition temperatures may be anticipated for borides due to boron's light mass and strong B-B covalent bonding, which yields high vibrational frequencies. Superconductivity has been discovered in transition metal diborides like MgB$_2$ (\textit{T}$_c$ = 39 K\cite{MgB2_X}), (Mo$_{0.96}$Zr$_{0.04}$)$_{0.85}$B$_2$ (\textit{T}$_c$ = 8.2 K\cite{MoZrB2_SC}), and NbB$_2$ (\textit{T}$_c$ = 5.2 K\cite{Akimitsu2001}). Often the superconducting early transition metal ``diborides'' are deficient in metal content, such as is seen in Nb$_{0.76}$B$_2$, which is deficient in Nb.\cite{NbB2_4, NbB2_5, Nunes} The metal deficiencies can sometimes be described by the general formula \textit{T}$_{1-\delta}$B$_{2+3\delta}$ where every time a transition metal (\textit{T}) is removed, 3 B atoms are added to the structure. Moreover, in these metal borides, boron atoms are likely to bond strongly with each other and form various molecule-like boron clusters, such as in the honeycomb lattice sheets of boron in MgB$_2$. 
	
	The strong B-B covalent bonding is not only attractive in superconducting materials but also in superhard materials, which have been extensively studied, especially for their practical use as cutting tools, abrasives, or wear-resistant coatings in various industrial applications.\cite{design_hardmat,MoB2_WB2,WB4_88,Rheniumdiboride_hard, B6O_hard,ultra_incompressible,TcB4,metalborides,CN_hard} Diamond and a variety of light-element compounds, although superhard materials, have practical and synthetic limitations for some applications\cite{diamond2,diamond1,cubic_BN,BC2N_hard} motivating the search for other types of superhard materials\cite{OsB2_hard1} like metal borides, which are an important class of superhard materials that can be easily synthesized. The well-known superhard metal diborides are RuB$_2$,\cite{OsB2_ReB2} OsB$_2$,\cite{OsB2_hard1,OsB2_hardmaterial} and ReB$_2$.\cite{ReB2_firstprinciples,ReB2_hard4} The shared characteristics of metal borides that link superconductors and superhard materials motivated us to study the superconducting properties of superhard materials. 
	
	``WB$_4$'', the subject of the current work, is also a superhard material, with a hardness of 43 GPa.\cite{WB2_hardness_scale} Structurally, WB$_4$ has been of particular interest because most other transition metals besides tungsten cannot tolerate higher boron contents and therefore exist as diborides.\cite{WB4_xx,WB4_cryst23} Most studies since the discovery of WB$_4$\cite{WB4_original} have found the crystal structure to have hexagonal symmetry, but there are a range of reported formulas such as WB$_4$,\cite{WB4_cryst23} WB$_{4.92}$,\cite{WB4.92} W$_{1-x}$B$_3$,\cite{W1-xB3} or WB$_{4.2}$\cite{PNAS_Kaner}.  The lattermost case is most consistent with our work. 
	
	Here we report the crystal structure and basic superconducting properties of the superhard boride WB$_{4.2}$. Our single crystal X-ray diffraction characterization shows, in agreement with earlier work, that WB$_{4.2}$ crystallizes in the space group \textit{P}6$_3$/\textit{mmc} (No. 194) and has a crystal structure that is derived from the simple diborides but with a systematic W-deficiency-B3 substitution - each missing W atom in the 2\textit{b} position is replaced by 3 B atoms in the 6\textit{h} position. Different experiments show sharp and reproducible superconducting transitions at about 2 K for WB$_{4.2}$. Temperature-dependent magnetic susceptibility, electrical resistivity and specific heat measurements were used to characterize the superconductor. WB$_{4.2}$ is a BCS-type superconductor with an upper critical magnetic field, \textit{H}$_{c2}$(0), that is between that predicted by the standard WHH equation and the result of using a linear extrapolation of \textit{H}$_{c2}$(\textit{T}) to zero temperature.

\section{Experimental Methods}
 
	The starting materials for the synthesis of polycrystalline WB$_{4.2}$ were boron (99.5$\%$, chunk, Johnson Matthey Catalog Co.) and tungsten powder ($>$99.9$\%$, Alfa). The tungsten powder was pressed into a pellet and arc-melted to have a metal chunk for subsequent meltings. The W and B chunks were weighed out in a 1:10 ratio and arc-melted three times in a Zr-gettered Ar atmosphere of $\sim$600 mbar. The arc-melted buttons were flipped in between each melting to ensure homogeneous samples. The resulting sample buttons had $<$1$\%$ mass loss and are stable in air over time. High quality samples of WB$_{4.2}$ are only obtained when a significant excess of B is employed, as has been previously reported.\cite{PNAS_Kaner} The purity of all samples was checked at room temperature using a Bruker D8 Advance Eco Cu K$_\alpha$ radiation ($\lambda$ =1.5406 $\textsc{\AA}$) X-ray diffractometer equipped with a LynxEye-XE detector. Single crystals from arc-melted samples were mounted on the tips of Kapton loops and room-temperature intensity data were collected using a Bruker Apex II X-ray diffractometer with Mo K$_{\alpha1}$ radiation ($\lambda$=0.71073 $\textsc{\AA}$). All data were collected with 0.5$^{\circ}$ scans in $\omega$ over a full sphere of reciprocal space with 10 s per frame of exposure time. Data acquisition was carried out using the SMART software.  The program VESTA was used to create all crystal structure images.\cite{VESTA} The SAINT program was used to both extract and correct intensities for polarization and Lorentz effects. XPREP, which is based on face-indexed absorption, was used for numerical absorption corrections.\cite{SHELXTL} The unit cell was tested for twinning. The crystal structure was then solved using direct methods with the SHELXTL package and the refinement was carried out by full-matrix least squares on F$^2$.\cite{SHELX} A LeBail fit of the room-temperature powder X-ray diffraction data was performed using the FullProf Suite program with Thompson-Cox-Hastings pseudo-Voigt peak shapes and the lattice parameters from the single crystal refinement as a starting point.
	 
	A Superconducting Quantum Interference Device with a $^3$He attachment was used to measure the zero-field cooled (ZFC) temperature-dependent volume magnetic susceptibility from 0.5 K to 2.2 K with \textit{H} = 20 Oe as the applied magnetic field. The field-dependent magnetization was measured at 1.4 K, 1.0 K, and 0.5 K from 0 - 35 Oe. A Physical Property Measurement System (PPMS) equipped with a resistivity option and $^3$He attachment was used to measure the temperature-dependent electrical resistivity from 300 K - 0.5 K under zero applied magnetic field using a standard four probe method. The low temperature resistivity was measured under applied magnetic fields ranging from 0 mT - 600 mT in the temperature region from 0.5 K - 2.2 K. The specific heat was measured on a small polished sample of polycrystalline WB$_{4.2}$ using Apiezon N grease from 3.6 K - 0.5 K under zero applied magnetic field. 

\section{Results and Discussion}
\begin{figure}[tbh!]
\includegraphics[scale = 0.38]{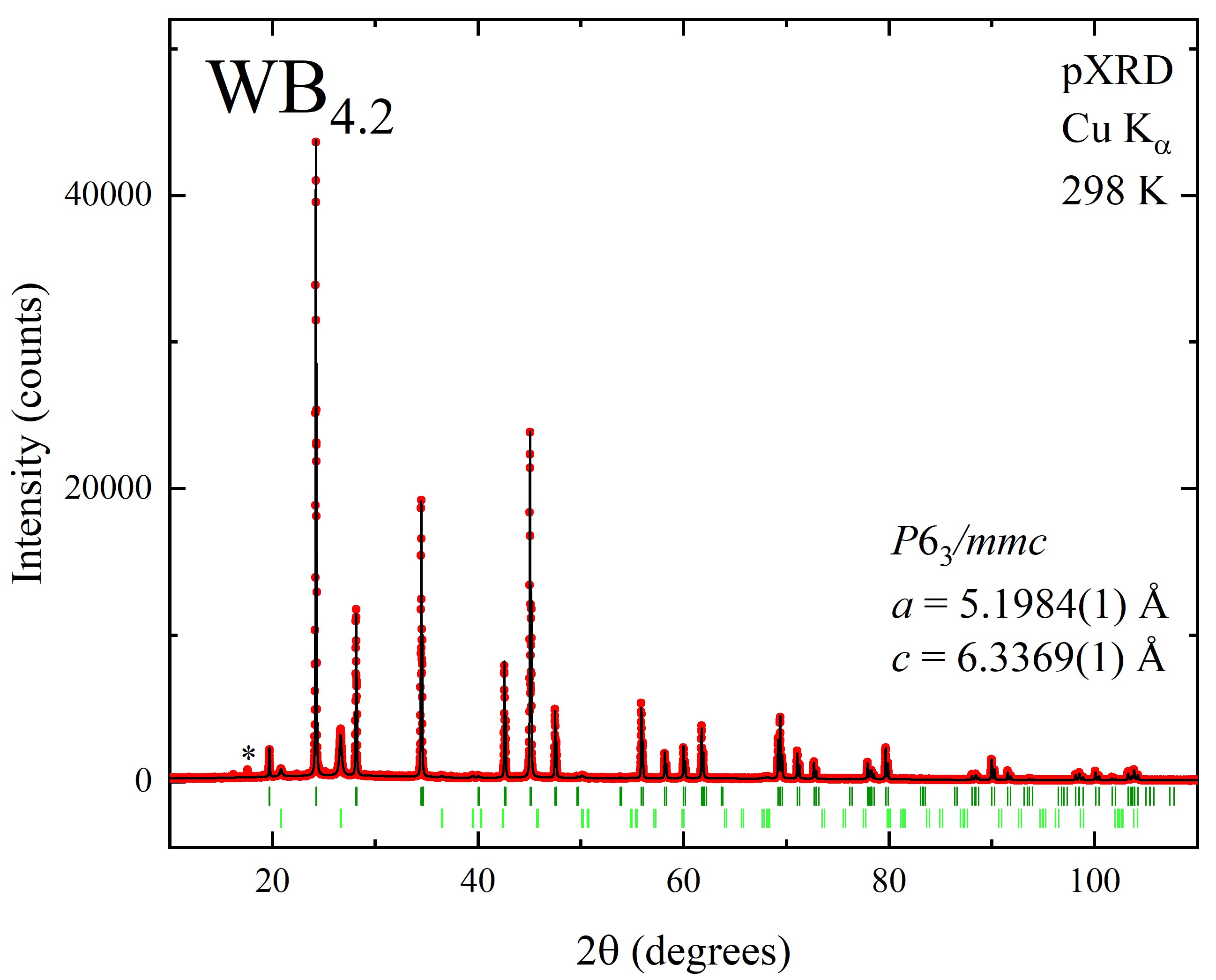}
\caption{LeBail fit of room temperature powder X-ray diffraction data for WB$_{4.2}$ indexed to space group \textit{P}6$_3$/\textit{mmc}. The red circles are the observed data, the black line is the calculated fit, the dark green vertical bars are the expected Bragg reflections for space group \textit{P}6$_3$/\textit{mmc}, and the light green vertical bars are the expected Bragg reflections for SiO$_2$ (from the mortar and pestle). The asterisk indicates a small amount of boron impurity.}
\label{Fig1_10}
\end{figure}

\begin{table}
\caption{Single crystal crystallographic data for WB$_{4.2(1)}$ at 293(2) K.}
\begin{tabular}{c c} % The final bracket specifies the number of columns in the table along with left and right borders which are specified using vertical bars (|); each column can be left, right or center-justified using l, r or c. To specify a precise width, use p{width}, e.g. p{5cm}
\hline 
Chemical Formula & WB$_{4.2(1)}$  \\
\hline  
F.W. (g/mol) 				& 229.33 							\\  
Space group 			& $P$6$_3$/$mmc$ (No. 194)		\\  
$a$ (${\textsc{\AA}}$)  	& 5.191(6)							\\ 
$c$ (${\textsc{\AA}}$)   	& 6.345(8)						\\

$V$ (${\textsc{\AA}^3}$) 	& 148.0(4)						\\
Absorption Correction	 	& Numerical							\\
Extinction Coefficient		& 	0.014(2)				\\
No. reflections; $R_{int}$	& 874; 0.0833						\\
No. independent reflections & 		151						\\
No. parameters				& 		15						\\
$R_1$; $wR_2$ (I$>$2$\sigma$(I))	& 0.0327; 0.0573		\\
$R_1$; $wR_2$ (all I)		&0.0556; 0.0606					\\
Goodness of fit				& 	1.130						\\
Diffraction peak and hole (e$^-$/${\textsc{\AA}^3}$)	& 2.640; -3.284\\
\hline% In-table horizontal line
\hline
\end{tabular}
\label{Table1_10} 
\end{table}

\begin{figure}[t]
\includegraphics[scale = 0.5]{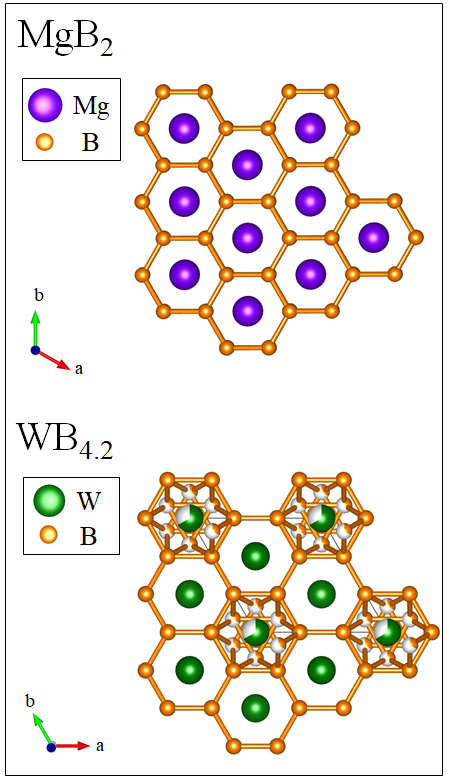}
\caption{Comparison of the crystal structures of MgB$_2$ (top) and WB$_{4.2}$ (bottom). Boron atoms are shown in orange, magnesium atoms in purple, and tungsten in green.}
\label{Fig2_10}
\end{figure}

	The crystal structure of WB$_{4.2}$ was analyzed using single crystal X-ray diffraction, which showed that the superconductor crystallizes in the space group \textit{P}6$_3$/\textit{mmc} (No. 194) with lattice parameters \textit{a} = 5.191(6) $\textsc{\AA}$ and \textit{c} = 6.345(8) $\textsc{\AA}$, consistent with what has been previously reported.\cite{PNAS_Kaner} Fig. \ref{Fig1_10} shows a LeBail fit of the room temperature powder X-ray diffraction (pXRD) data for WB$_{4.2}$ using the lattice parameters and the space group from the single crystal refinement as a starting point. It should be noted that there is some SiO$_2$ in the pXRD pattern (indicated by the light green vertical bars in Fig. \ref{Fig1_10}) that originates from parts of the agate mortar and pestle that intrude during grinding due to the superhardness of WB$_{4.2}$.  The lattice parameters from both refinements are in good agreement with one another so only the single crystal refinement will be discussed here. The results from the single crystal refinement are shown in Table \ref{Table1_10} and the atomic coordinates from the same refinement are given in Table \ref{Table2_10} for WB$_{4.2}$. The W2 atom position was freely refined, resulting in an occupancy of  0.665, and, consistent with the previous report, three B atoms were found in a triangular arrangement in place of the missing W2 atoms.\cite{PNAS_Kaner} Fig. \ref{Fig2_10} shows a comparison of the crystal structure of WB$_{4.2}$ (bottom) and MgB$_2$ (top). The boron atoms in MgB$_2$ form a honeycomb network, similar to graphite, with the magnesium atoms between the upper and lower six-membered rings that form the hexagonal honeycomb layers. WB$_{4.2}$ has a similar boron honeycomb with some of the holes between layers fully occupied by only W atoms while other holes can be occupied either by W2 or a B3 triangle. Although the occupancy of 0.665 for the W2 position is within error of a value of 2/3, which has the potential to result in a long range ordered structure, no evidence for W2 ordering is observed, consistent with previous reports.\cite{PNAS_Kaner}
 
\begin{table}
\caption{Atomic coordinates and equivalent isotropic displacement parameters of WB$_{4.2(1)}$ at 293(2) K. U$_{eq}$ is defined as one-third of the trace of the orthogonalized U$_{ij}$ tensor (${\textsc{\AA}^2}$).}
\begin{tabular}{ccccccc} % The final bracket specifies the number of columns in the table along with left and right borders which are specified using vertical bars (|); each column can be left, right or center-justified using l, r or c. To specify a precise width, use p{width}, e.g. p{5cm}
\hline
\hline % Top horizontal line
Atom & Wyckoff & Occ. & $\textit{x}$ & $\textit{y}$ & $\textit{z}$ & U$_{eq}$\\ % Column names row
\hline % In-table horizontal line
W1	& 2\textit{c} 		& 1 	& 2/3				 & 1/3      	& 1/4& 0.0029(3) \\
W2	& 2\textit{b} 		& 0.665(8)		 & 0				 & 0      	& 1/4 &0.0038(4) \\
B1 	& 12\textit{i} 		& 1 & 0.335(4) 		& 0 &0 &0.005(2)\\  
B2 	& 6\textit{h} 		& 0.335$^\#$ & 0.115(5) 		& 0.229(9) &1/4 &0.008(3)\\

\hline% In-table horizontal line
\hline % Bottom horizontal line
\end{tabular}
\label{Table2_10} 
\begin{flushleft}
$^\#$The W2 position occupancy is freely refined, but the fractional occupancy of the B2 site is constrained such that the occupancy of B2 = 1 - occupancy of W2, as described in previous studies.\cite{PNAS_Kaner} The positional parameters of B2 are freely refined.
\end{flushleft}
\end{table}

\begin{figure}[t]
\includegraphics[scale = 0.35]{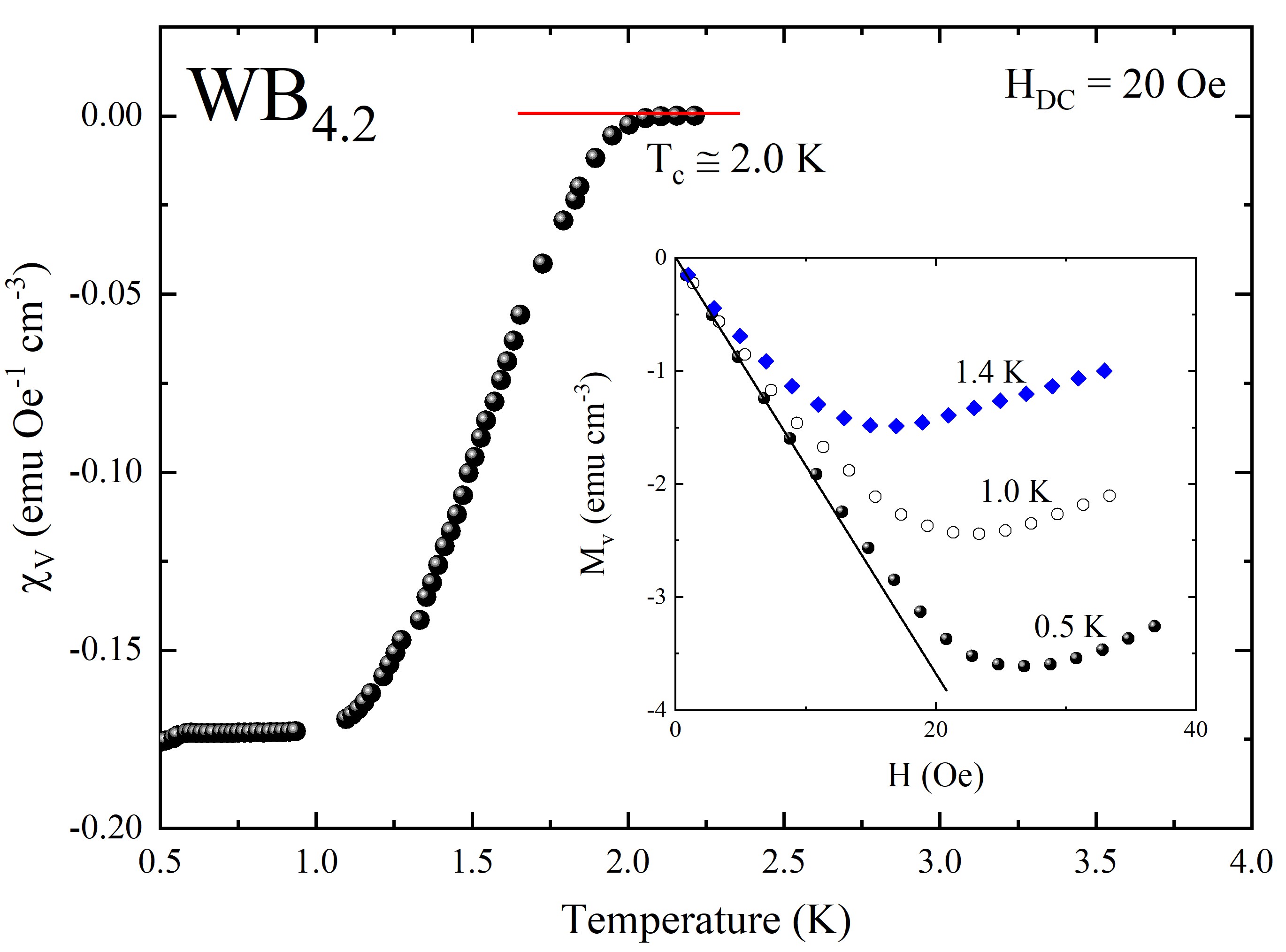}
\caption{Temperature dependence of zero-field-cooling (ZFC) volume magnetic susceptibility ($\chi_V$) measured in a 20 Oe applied magnetic field. Inset: Volume magnetization (\textit{M}$_V$) vs. field at 0.5 K, 1.0 K and 1.4 K.}
\label{Fig3_10}
\end{figure}

The temperature-dependent volume magnetic susceptibility ($\chi_V$) was measured to characterize the superconducting critical temperature of WB$_{4.2}$. The zero-field cooled (ZFC) $\chi_V$ vs \textit{T} (Fig. \ref{Fig3_10}) shows the first clear deviation from the normal state susceptibility at approximately 2.0 K. The volume magnetization (\textit{M}$_V$) was measured at 1.4 K, 1.0 K, and 0.5 K from 0 to 35 Oe as shown in the inset of Fig. \ref{Fig3_10}. A rough estimation of the lower critical field at 0.5 K (not corrected for demagnetization) is \textit{H}$_{c1}$(0.5 K) = 10 Oe.

\begin{figure}[t]
\includegraphics[scale = 0.35]{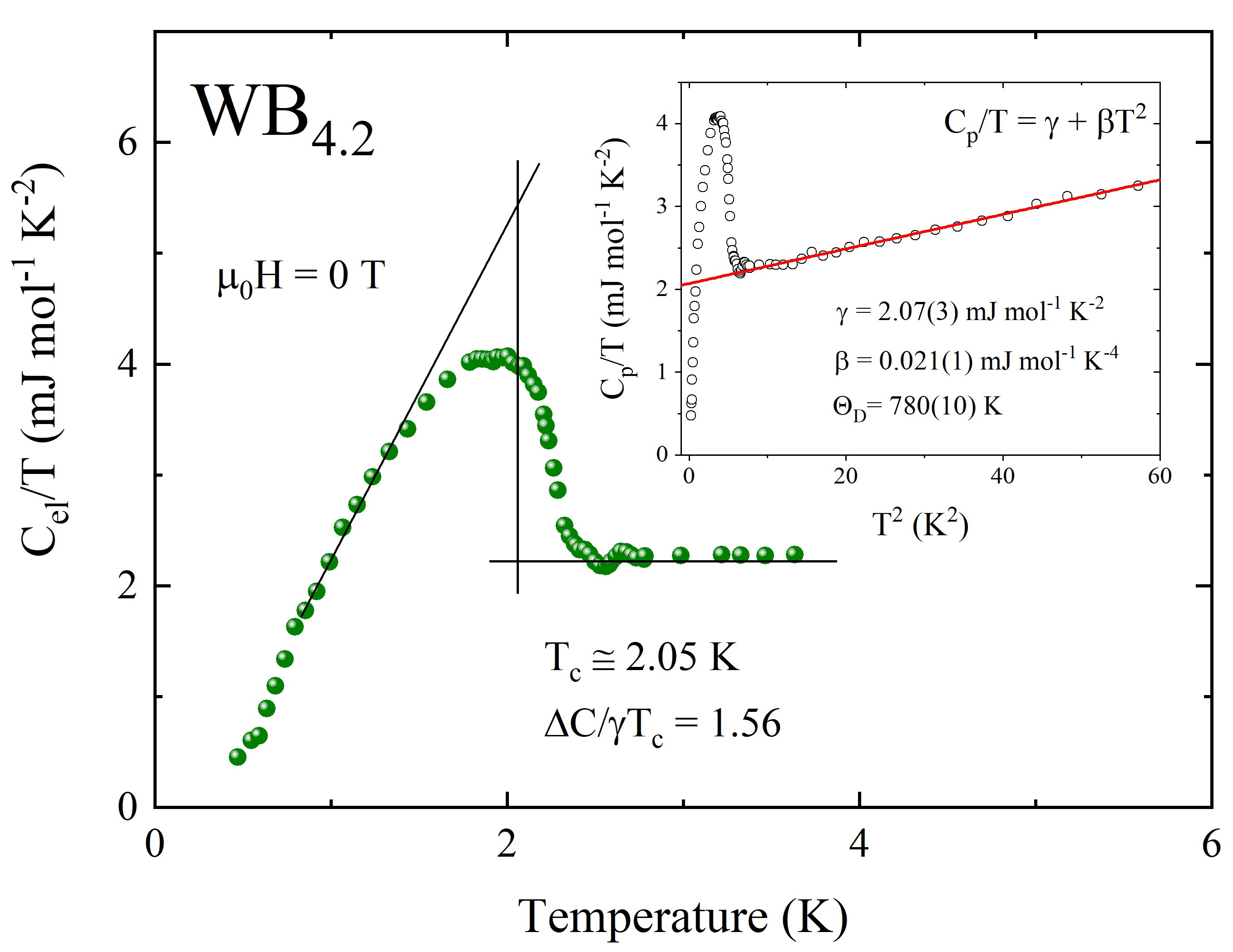}
\caption{Electronic heat capacity divided by temperature (\textit{C}$_{el}$/\textit{T}) vs. temperature measured from 3.6 K to 0.5 K under zero applied magnetic field. Inset: \textit{C}$_p$/\textit{T} vs. \textit{T}$^2$ shown for the low temperature region and fitted to a line.}
\label{Fig4_10}
\end{figure}

The inset of Fig. \ref{Fig4_10} shows \textit{C}$_p$/\textit{T} vs \textit{T}$^2$, where the data were fitted above the critical temperature to the equation,
\begin{equation}
\frac{C_p}{T} = \gamma + \beta T^2
\label{Eq2}
\end{equation}
\noindent where $\gamma$\textit{T} is the electronic contribution (\textit{C}$_{el}$) to the specific heat and $\beta$T$^3$ is the phonon contribution (\textit{C}$_{ph}$). The slope of the fitted line, $\beta$, is 0.021(1) mJ mol$^{-1}$ K$^{-4}$ and the Sommerfeld parameter, $\gamma$, is calculated to be 2.07(3) mJ mol$^{-1}$ K$^{-2}$. Using $\beta$, the Debye temperature ($\Theta_D$) is calculated to be 780(10) K using the equation 
\begin{equation}
\Theta_D = \left(\frac{12\pi^4}{5\beta} n R\right)^\frac{1}{3}
\label{Eq3}
\end{equation}
\noindent where \textit{R} is the ideal gas constant 8.314 J mol$^{-1}$ K$^{-1}$ and \textit{n} = 5.2 for WB$_{4.2}$. The Debye temperature is high, which reflects a high concentration of boron (for which $\Theta_D$ = 1300 K\cite{lattice_B}) in the material. With $\Theta_D$, \textit{T}$_c$, and assuming $\mu^*$= 0.13, the inverted McMillan formula\cite{McMillan}
\begin{equation}
\lambda_{ep} = \frac{1.04 + \mu^*\ln\left( \frac{\Theta_D}{1.45T_c}\right)}{(1-0.62\mu^*)\ln\left( \frac{\Theta_D}{1.45 T_c}\right) - 1.04}
\label{Eq4}
\end{equation}
\noindent can be used to calculate the electron-phonon coupling constant $\lambda_{ep}$ to be 0.43 for WB$_{4.2}$. The Fermi energy N(E$_F$) is calculated with the formula 
\begin{equation}
N(E_F) = \frac{3\gamma}{\pi^2 k_B^2(1 + \lambda_{ep})},
\label{Eq5}
\end{equation}
\noindent to be 0.61 states eV$^{-1}$ per formula unit of WB$_{4.2}$ where $\gamma$ = 2.07(3) mJ mol$^{-1}$ K$^{-2}$, $\lambda_{ep}$ = 0.43, and k$_B$ is the Boltzmann constant.

The main panel of Fig. \ref{Fig4_10} shows the electronic specific heat divided by temperature (\textit{C}$_{el}$/\textit{T}) vs temperature for WB$_{4.2}$ from 0.5 K - 3.6 K with zero applied magnetic field showing a large peak in the specific heat. \textit{C}$_{el}$ was obtained by subtracting the phonon contribution to the specific heat: \textit{C}$_{el}$ = \textit{C}$_p$ - $\beta$\textit{T}$^3$.  The \textit{T}$_c$ is estimated to be 2.05 K for this data using an equal-area entropy construction (solid black lines), which is close to the critical temperature from the temperature-dependent magnetic susceptibility measurement. Such a large peak in the specific heat due to a significant loss of entropy is an explicit indication that the bulk material is superconducting. The normalized specific heat jump, $\Delta$\textit{C}/$\gamma$\textit{T}$_c$, is 1.56 for WB$_{4.2}$, which is slightly larger than the expected weak coupling BCS limit of 1.43\cite{intro_SC} and confirms bulk superconductivity in our material. 

\begin{figure}[t]
\includegraphics[scale = 0.35]{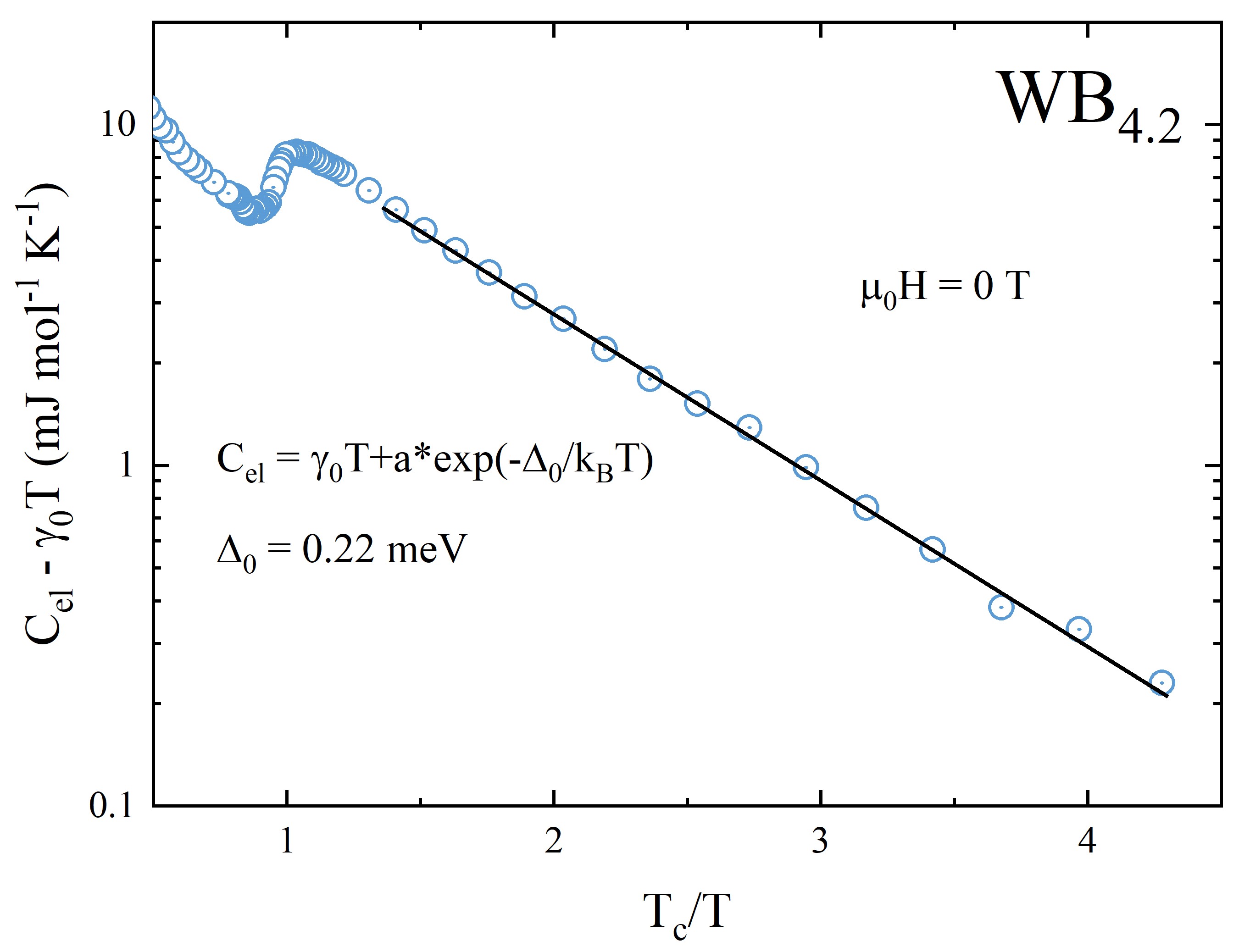}
\caption{Electronic heat capacity \textit{C}$_{el}$.-$\gamma_0$\textit{T} vs. the normalized \textit{T}$_c$/\textit{T} for WB$_{4.2}$ to determine the superconducting gap $\Delta_0$.}
\label{Fig5_10}
\end{figure}

\begin{figure}[b]
\includegraphics[scale = 0.35]{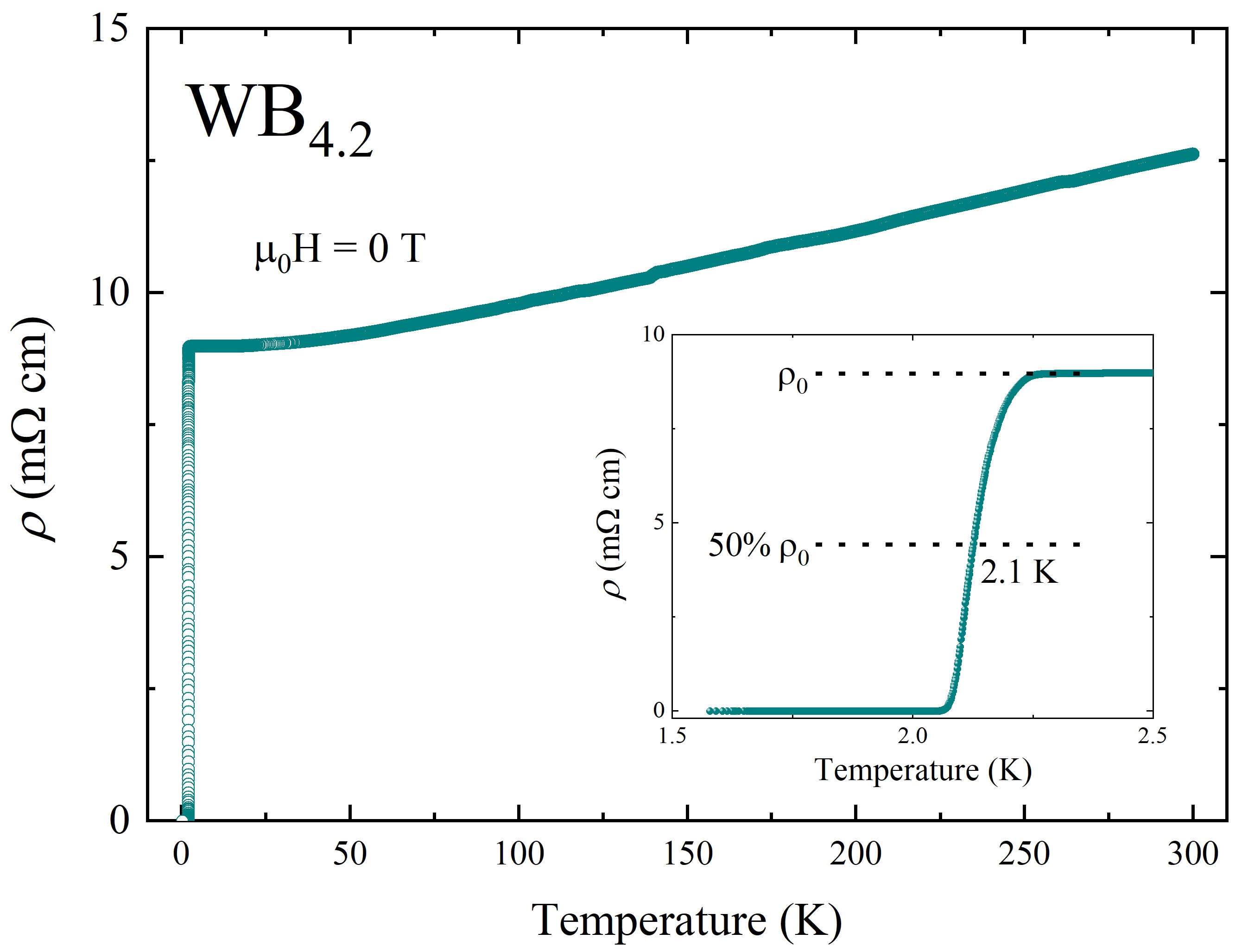}
\caption{Temperature-dependent resistivity of polycrystalline WB$_{4.2}$ measured from 300 K - 1.7 K under zero applied magnetic field. Inset: Resistivity near the superconducting transition.}
\label{Fig6_10}
\end{figure}

Fig. \ref{Fig5_10} shows \textit{C}$_{el}$-$\gamma_0$\textit{T} vs \textit{T}$_c$/\textit{T} for WB$_{4.2}$ in the superconducting state under zero applied magnetic field and fitted to the following equation 
\begin{equation}
C_{el} = \gamma_0T + a*e^\frac{-\Delta_0}{k_BT},
\label{Eq5}
\end{equation}
\noindent where $\gamma_0$\textit{T} is the electronic contribution to the specific heat originating from impurities in the sample, $\Delta_0$ is the superconducting gap magnitude, and k$_B$ is the Boltzmann constant. The $\Delta_0$ for  WB$_{4.2}$ was calculated to be $\Delta_0$ = 0.22 meV and according to BCS theory,\cite{intro_SC} it is expected to be 
\begin{equation}
2\Delta_0 = 3.5k_BT_c
\label{Eq5}
\end{equation}
\noindent for a weak coupling superconductor. The calculated value of the superconducting gap $\Delta_0$ = 0.22 meV (2$\Delta_0$ = 0.44 meV) is less than the weak coupling BCS value of 2$\Delta_0$ = 0.62 meV.

\begin{figure}[b]
\includegraphics[scale = 0.35]{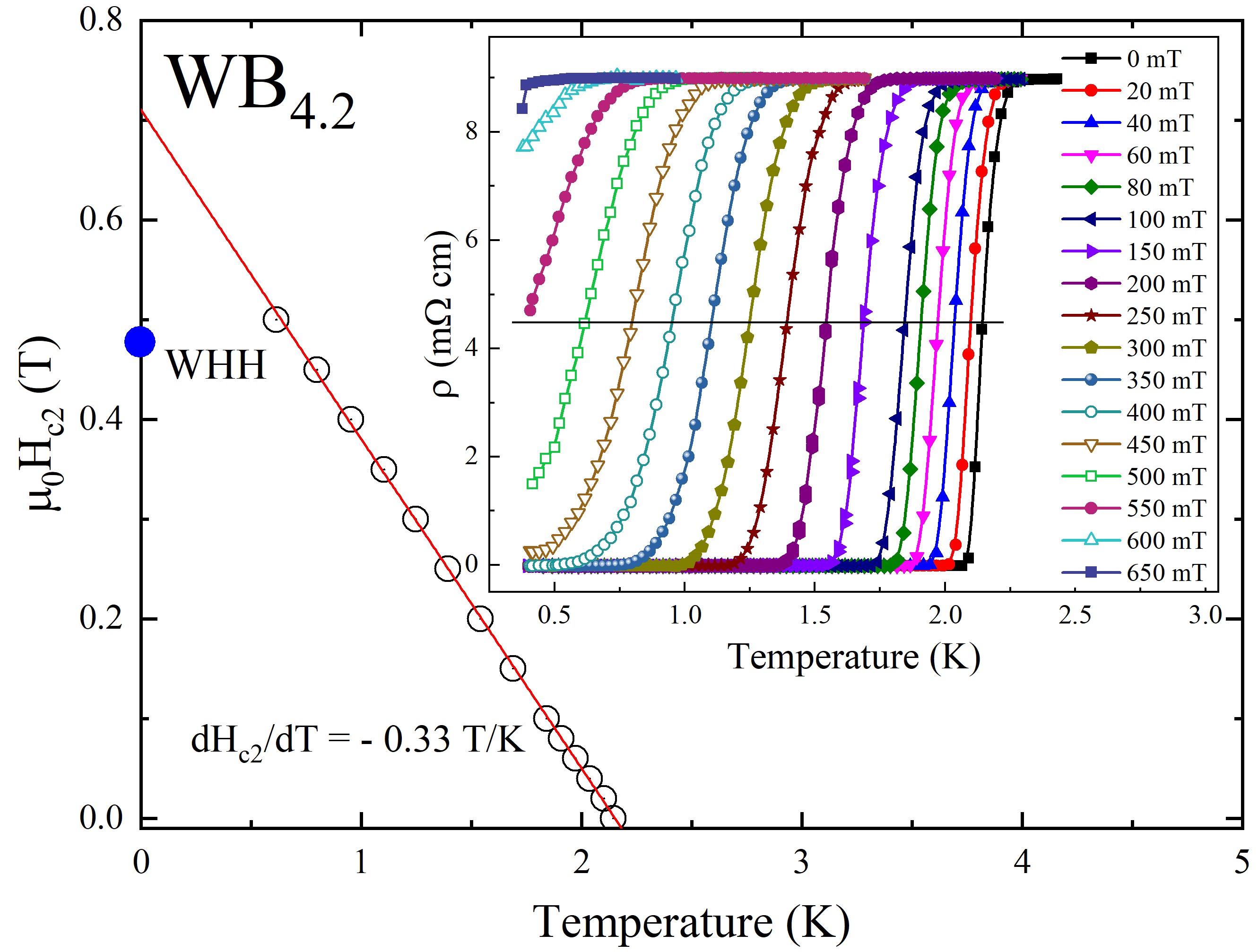}
\caption{Upper critical field ($\mu_0$\textit{H}$_{c2}$) vs temperature fitted to a line. Inset: $\rho$ vs. \textit{T} under various magnetic fields showing the dependence of the \textit{T}$_c$ on the applied field. The solid black line shows 50$\%$ of the superconducting transition.}
\label{Fig7_10}
\end{figure}

Fig. \ref{Fig6_10} shows the temperature-dependent electrical resistivity for WB$_{4.2}$ measured from 300 K - 1.7 K. In the normal state, the resistivity changes slightly and the RRR ratio ($\rho_{300K}$/$\rho_{2.2K}$) is only $\sim$1.3, as is often observed for polycrystalline samples. The inset of Fig. \ref{Fig6_10} shows that the resistivity drops to 50 $\%$ of the normal state value at $\sim$ 2.1 K (black dotted line) and reaches zero above 2 K. The resistivity transition is sharp and only covers a 0.2 K temperature range. The \textit{T}$_c$ values from the resistivity, specific heat, and magnetic susceptibility are in agreement. The inset of Fig. \ref{Fig7_10} shows the dependence of the \textit{T}$_c$ (black line) on the applied magnetic field for WB$_{4.2}$ in the temperature range from 0.5 K - 2.2 K under applied magnetic fields ranging from 0 mT - 600 mT. The \textit{T}$_c$ steadily decreases with increasing applied field, as expected, and the last instance where the resistivity drops to below 50 $\%$ is for \textit{T}$_c$ = 0.62 K with $\mu_0$\textit{H} = 500 mT. The thus-determined upper critical fields ($\mu_0$\textit{H}$_{c2}$) plotted as a function of the estimated \textit{T}$_c$ values were plotted in the main panel of Fig. \ref{Fig7_10} and fitted to a line with slope d$\mu_0$\textit{H}$_{c2}$/d\textit{T} = -0.33 T/K. For many superconductors, the zero temperature upper critical field $\mu_0$\textit{H}$_{c2}$(0) can be estimated with the Werthamer-Helfand-Hohenberg (WHH)\cite{cleanlimit} equation given by: 
\begin{equation}
\mu_0H_{c2}(0) = -A T_c \frac{d\mu_0H_{c2}}{dT}\bigg|_{T=T_c},
\label{Eq6}
\end{equation}
where \textit{A} is -0.693 for the dirty limit and taking \textit{T}$_c$ as $\sim$2.05 K for WB$_{4.2}$. Based on this model, the $\mu_0$\textit{H}$_{c2}$(0) = 0.47 T (indicated by the blue closed circle in Fig. \ref{Fig7_10}), however the last measured $\mu_0$\textit{H}$_{c2}$ value is 0.50 T which is already above the WHH-predicted upper critical field of 0.47 T. The nearly linear \textit{H}$_{c2}$(\textit{T}) over a broad temperature range that we observe has been seen previously for Fe based superconductors\cite{BaFeCoAs2,BaKFe2As2} and also for Nb$_2$Pd$_{0.81}$S$_5$\cite{Nb2Pd0.81S5} and claimed to originate from the multi-band superconductivity effect. The resulting $\mu_0$\textit{H}$_{c2}$(0) for WB$_{4.2}$ is therefore most likely somewhere between the linear extrapolation to 0 K (0.71 T) and that predicted by the WHH model (0.47 T). Taking $\mu_0$\textit{H}$_{c2}$(0) = 0.71 T as the upper limit, the approximate coherence length can be calculated by using the Ginzburg - Landau formula $\xi_{GL}$(0) = ($\Phi_0$/[2$\pi$\textit{H}$_{c2}$(0)])$^{1/2}$, where $\Phi_0$ = \textit{h}/2\textit{e} and is found to be $\xi_{GL}$(0) = 26 nm. All physical parameters for WB$_{4.2}$ are gathered in Table \ref{Table3_10}.

\begin{table}
\caption{Normal and superconducting state parameters for WB$_{4.2}$.}
\begin{tabular}{p{3cm}p{3cm}p{2.2cm}} % The final bracket specifies the number of columns in the table along with left and right borders which are specified using vertical bars (|); each column can be left, right or center-justified using l, r or c. To specify a precise width, use p{width}, e.g. p{5cm}

\hline % Top horizontal line
Parameter & Units & Value \\% Column names row

\hline % In-table horizontal line
\textit{T}$_c$						& K		&  2.05(5) \\  
$\mu_0$\textit{H}$_{c2}$(0) 			& T		&  0.71\\ 
$\lambda_{ep}$ 			& -	&  0.43\\
$\xi_{GL}(0)$ 			& nm	&  26\\ 
$\gamma$ 			& mJ mol$^{-1}$K$^{-2}$	&  2.07(3)\\
$\beta$				&mJ mol$^{-1}$K$^{-4}$	&0.021(1)\\
$\Theta_{D}$				& K				& 780(10)\\
$\Delta$C/$\gamma$\textit{T}$_c$ 			& -	&  1.56\\
$\Delta_0$			& meV	&0.22\\
N(E$_F$)			&  states eV$^{-1}$ f.u.$^{-1}$	&  0.61\\

\hline% In-table horizontal line
\hline % Bottom horizontal line
\end{tabular}
\label{Table3_10} 
\end{table}

\section{Conclusions}

We report superconductivity in the superhard material WB$_{4.2}$ and confirm that this material crystallizes in the hexagonal space group \textit{P}6$_3$/\textit{mmc} using room-temperature single crystal X-ray diffraction data. This material is non-stoichiometric via W deficiency, where for every missing W atom, 3 B atoms are inserted into the structural cavity in its place. The superconducting transition occurs at about 2 K and is characterized through temperature-dependent magnetic susceptibility, electrical resistivity, and specific heat measurements. WB$_{4.2}$ is shown to be a BCS-type weak coupled superconductor based on the calculated superconducting parameters. The $\mu_0$\textit{H}$_{c2}$(0) value is between the value predicted using the WHH model and that of a linear extrapolation of $\mu_0$\textit{H}$_{c2}$(\textit{T}) to 0 K. This material appears to be an example of a highly nonstoichiometric, metal deficient, early transition metal diboride superconductor.

\section*{Acknowledgements}

The materials synthesis was supported by the Department of Energy, Division of Basic Energy Sciences, Grant DE-FG02-98ER45706, and the property characterization at Princeton was supported by the Gordon and Betty Moore Foundation EPiQS initiative, Grant GBMF-4412. The research at the Gdansk University of Technology was supported by the National Science Centre, Grant UMO-2016/22/M/ST5/00435. The work in the Department of Chemistry at LSU was supported by the US Department of Energy under EPSCoR Grant No. DE-SC0012432 with additional support from the Louisiana Board of Regents.    

\section*{Author Correspondence}
\begin{flushleft}
$^*$E.M.C (carnicom@princeton.edu)\\
$^*$R.J.C. (rcava@exchange.princeton.edu)
\end{flushleft}

\bibliographystyle{apsrev4-1}
\bibliography{MyBib}

\end{document}